\documentclass[aps,pra,showpacs,twocolumn,superscriptaddress,preprintnumbers,amsmath,amssymb]{revtex4}

\usepackage{graphicx}
\usepackage{bm}

\usepackage[usenames]{color}
\usepackage{textcomp}
\begin{document}

\title{Lattice-Ramp Induced Dynamics in an Interacting Bose-Bose Mixture}
\author{Julia Wernsdorfer}
\affiliation{Institut f\"ur Theoretische Physik, Johann Wolfgang 
Goethe-Universit\"at, 60438 Frankfurt/Main, Germany}
\author{Michiel Snoek}
\affiliation{Instituut voor Theoretische Fysica, Universiteit van Amsterdam, 1018 XE Amsterdam, Netherlands}
\author{Walter Hofstetter}
\affiliation{Institut f\"ur Theoretische Physik, Johann Wolfgang 
Goethe-Universit\"at, 60438 Frankfurt/Main, Germany}
\date{\today}

\begin{abstract}
We investigate a bosonic quantum gas consisting of two interacting species in an optical lattice at zero and finite temperature. The equilibrium properties and dynamics of this system are obtained by means of the Gutzwiller mean-field method. In particular we model recent experiments where the ramp-up of the optical lattice occurs on a time scale comparable to the tunneling time of the bosons. We demonstrate the violation of adiabaticity of this process with respect to the many-body quantum states, and reproduce and explain the oscillations of the visibility as a function of ramp-up time, as seen in experiments. 

\end{abstract}


\maketitle
\section{Introduction}
Ultracold gases provide a powerful system for the experimental investigation of interacting quantum many-body systems. In combination with optical lattices and tunable inter- and intraspecies interactions, degenerate quantum gases give insight into both strongly and weakly correlated regimes, and possible phase transitions. They allow the direct probing of fascinating phenomena like the superfluid to Mott-insulator (SF-MI) transition. In the breakthrough experiment \cite{greiner} this transition was shown for the first time, realizing predictions for the well-known Bose-Hubbard model \cite{jaksch,fisher} in the case of a single bosonic species.

Adding a second atomic species results in a wealth of quantum phases, clearly demonstrating the complexity of correlated ensembles. Currently, several experimental groups are working on Fermi-Bose \cite{esslinger,best,fb-florenz}, Fermi-Fermi \cite{fermi-fermi,Ulrich,Esslinger-fermions,hansch} and Bose-Bose \cite{catani,bose-bose-Trotzky,ketterle} mixtures in optical lattices as they are promising devices for studies of disorder \cite{ospelkaus}, dipolar molecule formation \cite{damski}, and spin arrays \cite{kuklov}. In Bose-Fermi mixtures \cite{irakli_FB,buchler-blatter} and Bose-Bose mixtures \cite{hubener,soyler}, a supersolid phase has been predicted to exist. 

Recently the Florence group has realized a Bose-Bose mixture of $^{87}$\!Rb and $^{41}$\!K trapped in a 3D optical lattice \cite{catani}. They showed that the SF-MI transition of Rubidium is shifted towards shallower optical lattices when Potassium is present in the system. The same reduction of the visibility was seen in experiments where fermions were added to bosons in optical lattices \cite{esslinger, best,fb-florenz}.

These experiments are important steps towards the study of low-temperature properties of atomic mixtures.
The experimental investigation of quantum many-body ground-states, for instance with spin ordering, requires the achievement of rather low temperatures in the lattice. For fermions this already poses problems before ramping up the optical lattice, since evaporative cooling becomes inefficient in the degenerate limit. Bosons, on the other hand, can be cooled to very low temperatures ~\cite{temp-kett}. However, ramping up the optical lattice can easily increase the temperature, or more precisely the entropy, again. This poses the question we address in this paper: under which conditions is the ramping up of the optical lattice sufficiently slow, such that the process is adiabatic. This question has been investigated before for the case of a single species bosonic gas; both by the mean-field technique we employ here and by methods tailored to the one-dimensional systems \cite{dynamics2,buonsante,polen}. 

Here we systematically map out the visibility of a Bose-Bose mixture as a function of ramp-up time. We reproduce and explain the experimentally observed oscillations and relate them to the issue of adiabaticity. We take into account that this system breaks the rotational symmetry of the trap, because the center of masses of the two species are shifted with respect to each other due to the gravitational sag. We consider a two-dimensional system. In this way we keep the numerical effect manageable, and yet avoid the peculiarities associated with a one-dimensional system, like the absence of long range order even at zero temperature, which makes the extrapolation of one-dimensional results to higher dimensions problematic.

The timescale for ramping the optical lattice in the experiment is usually determined such that the system ends up in the lowest band of the optical lattice, i.e. the ramping time is chosen large with respect to the band gap. Since the band gap is small for shallow optical lattices, sometimes an exponential ramp-up profile is chosen. However, this does not guarantee that the ramping process is also adiabatic with respect to the many-body states in the lowest band \cite{polen, dynamics2}. This is indeed not true and can be easily seen from the fact that the tunneling time for the atoms $t_{\mathrm{hop}}(s)=\frac{\pi}{2}\frac{\hbar}{J_{Rb}(s)}$ is of the same order as the ramp-up time in most experiments. 

The results of our numerical simulations are in agreement with this qualitative argument. By simulating the ramp-up of the optical lattice using Gutzwiller mean-field theory, we observe that in the regime of deep optical lattices the ramp-up dynamics generally does not lead to the ground state of the system. The ground state of the system at those lattice depths typically contains a Mott insulating plateau with integer filling in the center. The states obtained after ramping up the lattice indeed lose the long-range superfluid order in this region. However, unlike the MI-phase, the particle densities are non-integer. We show that the exponential ramping profile is even less adiabatic than a linear increase of the depth of the optical lattice. 

By systematically tracing the visibility as a function of the ramp-up time, we find characteristic oscillations, which also have been observed in experiments with a single bosonic species \cite{bloch-osci} and a two-component bosonic mixture \cite{catani}. The fact that we reproduce these oscillations within the mean-field dynamics, where heating due to three-body collisions is not included, indicates that they are part of the real many-body dynamics.
We interpret these oscillations as a competition between two effects. The first effect is the finite time the particles need to localize. For fast ramps, the number of delocalized particles is far higher than the equilibrium value. However,  phase coherence is generally not present. The second effect is a coupling to collective excitations, in particular the breathing mode of the system. Whereas the first effect enhances the visibility, the second effect destroys it, because the induced current leads to occupation of finite momentum modes. The criterion for adiabaticity, we deduce from this, is the \textit{saturation of the visibility} as a function of ramp-up time. In this regime the superfluid fraction reaches its equilibrium value and the ramp-up is sufficiently slow such that no collective excitations are excited. The maximum of the visibility, as sometimes used as an experimental criterion, turns out to be not a good indication of adiabaticity.

This paper is organized as follows: in section ~\ref{sec:model} we present the two-component Bose-Hubbard model we investigate and explain the Gutzwiller mean-field method we apply. In section ~\ref{sec:resultsT0} we present results for $T=0$. In section ~\ref{sec:resultsT} we discuss effects of finite temperature. In section ~\ref{sec:conclusions} we present conclusions and implications for the experiments.

\section{Model}
\label{sec:model}

\begin{figure}
		\includegraphics[width=1.0\linewidth]{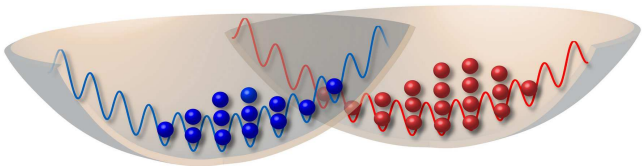}
	\caption{(Color online) Modeling of the gravitational sag by two spatially separated parabolic trapping potentials. The lattice potential with the superimposed harmonic trapping potential (here drawn in one dimension only) is represented by a sinusoidal blue and red line. Different colors represent the two different bosonic species.}
	\label{fig:Lattice_RbK_julia}
\end{figure}

The two-species bosonic system we consider is described by the Bose-Hubbard Hamiltonian:
\begin{eqnarray}
\hat{\mathcal{H}} &=& \sum_{\alpha=a,b}\big[\sum_{\langle ij \rangle}J_\alpha(\hat b^\dagger_{i\alpha}\hat b^{}_{j\alpha} + h.c.)+\sum_i \frac{U_\alpha}{2}\hat n_{i\alpha}(\hat n_{i\alpha}-1) \nonumber\\&& - \sum_i( \mu_{\alpha}-V^{trap}_{i\alpha})\hat n_{i\alpha}\big]+\sum_i U_{ab}\hat n_{ia}\hat n_{ib} \quad,
\label{eq:BH}
\end{eqnarray}
where the operator $\hat b^\dagger_{i\alpha} (\hat b^{}_{i\alpha})$ creates (annihilates) a boson of flavor $\alpha$, $\alpha=a, b$ at site $i$.  Since our goal is to model the experiment \cite{catani}, we choose the two species as $^{87}$Rb and $^{41}$K. The separation of the center of mass of the two clouds due to the gravitational sag is provided by two spatially separated parabolic traps $V^{trap}_{i\alpha} = \frac{1}{2}m_\alpha\omega^2_\alpha d^2| {\bf r}_i - {\bf r}_\alpha|^2$, where $d$ is the lattice constant, $\omega_\alpha$ the trap frequency, ${\bf r}_i$ are the coordinates of the given lattice site $i$ and ${\bf r}_\alpha$ are the centers of the harmonic potentials, see Fig.~\ref{fig:Lattice_RbK_julia}. The frequencies of the parabolic trap are $\omega_{Rb}=2\pi\times36\mathrm{Hz}$ and $\omega_K=2\pi\times53\mathrm{Hz}$. The distance between the two trap centers was chosen in such a way that the atomic clouds overlap only on a few lattice sites, as in the experiment \cite{catani}. 
The chemical potentials $\mu_{\alpha}$ are adjusted such that the resulting particle numbers correspond to the particle number
ratio of the experiment $N_{K}/N_{Rb}\approx0.1$ \cite{catani} ($N_{K}\approx 30, N_{Rb}\approx 300$). The parameters $U_{\alpha},U_{ab}, J_{\alpha}$ indicate the intra-/interspecies Coulomb repulsion and the hopping amplitude. As input we use the experimental values of the Florence experiments \cite{catani}: lattice laser wavelength $\lambda_L=1064$\,nm, scattering length $a_{Rb-K} = 163a_0$, $a_{Rb}=99a_0$, $a_{K}=65a_0$, with $a_0$ the Bohr radius. The hopping constants $J_\alpha$ and the interaction parameters $U_\alpha$, $U_{ab}$ are calculated according to \cite{bloch}:
\begin{eqnarray}
J_{\alpha}&=&\frac{4}{\sqrt{\pi}}\frac{\hbar^2}{2m_\alpha}\bigg(\frac{2\pi}{\lambda_L}\bigg)^2s^{3/4}_{\alpha}e^{-2\sqrt{s_{\alpha}}}\,,\\ \label{eq:param-formel}
U_{\alpha}&=&\sqrt{\frac{8}{\pi}}\frac{2\pi}{\lambda_L}a_{\alpha}E_r^{\alpha}s_{\alpha}^{3/4}\,,\\ 
U_{Rb-K}&=&\frac{4}{\sqrt{\pi}}ka_{Rb-K}E_r^{\alpha}\frac{1+\frac{m_{Rb}}{m_K}}{(1+\sqrt{\frac{m_{Rb}V^{Rb}_L}{m_{K}V^{K}_L}})^{3/2}}s_{Rb}^{\frac{3}{4}}
\end{eqnarray}
\begin{figure}[t!]
\includegraphics[width=1.0\linewidth]{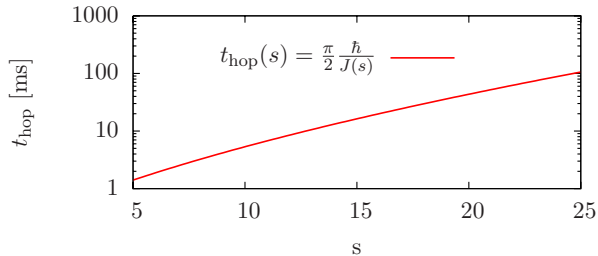}
\caption{The tunneling time as a function of the lattice depth.}
\label{fig:thop}
\end{figure}
with the dimensionless lattice depth $s_{\alpha}=V_L^{\alpha}/E_r^{\alpha}$, where $V_L^{\alpha}$ is the depth of the laser-induced potential and $E_r^{\alpha}$ the recoil energy of species $\alpha$. From the hopping constant we directly obtain the tunneling time of Rb-atoms $t_{\mathrm{hop}}(s)=\frac{\pi}{2}\frac{\hbar}{J_{Rb}(s)}$, which we plot as a function of lattice depth in Fig.~\ref{fig:thop}. For the given wavelength the depth of the laser induced lattice potential is species-specific: $V_L^{Rb}=1.1V_L^K$ and $E_r^{K}=2.1E_r^{Rb}$. This results in $s_{Rb}=2.3s_K$. In the following we use the short-hand notation $s=s_{Rb}$.
Here we neglect the influence of the parabolic confinement on the parameters. This is justified because experimentally the potential energy difference between neighboring lattice wells is much smaller than the barrier height \cite{formel}.

\subsection{Gutzwiller Mean-Field approximation}
The Hamiltonian Eq. (\ref{eq:BH}) can in principle be analyzed by Quantum Monte Carlo simulations \cite{QMCkashur,QMCgerbier}. However, these simulations are restricted to equilibrium situations and cannot describe dynamical processes like the ramping up of the optical lattice we consider here. Alternatively one can apply the time-evolving block decimation (TEBD) method to simulate the dynamics \cite{polen}. However, TEBD is restricted to one dimension.
Here we simulate the experimental ramp-up dynamics within a mean-field approximation, which provides good  qualitative results \cite{dynamics2}, requiring a much lower computational effort than exact numerical alternatives. 

Within the Gutzwiller method a mean-field approximation is applied to the operators $\hat b^\dagger_{i\alpha} (\hat b^{}_{i\alpha})$ in the hopping part of the Hamiltonian Eq. (\ref{eq:BH}). This leads to a decomposition of the lattice Hamiltonian into a sum over decoupled single-site Hamiltonians, which can be solved numerically:
\begin{eqnarray}
\hat{\mathcal{H}}^{MF}_i &=& \sum_{\alpha=a,b}\big[J_\alpha(\phi^*_{i\alpha}\hat b^{}_{i\alpha} + h.c.)+\frac{U_\alpha}{2}\hat n_{i\alpha}(\hat n_{i\alpha}-1) \nonumber\\&& - ( \mu_{\alpha}-V^{trap}_{i\alpha})\hat n_{i\alpha}\big]+ U_{ab}\hat n_{ia}\hat n_{ib} \quad,
\label{eq:BH-MF}
\end{eqnarray}
with $\phi_{i\alpha}=\sum_{j n.n. i}\langle\hat b^{}_{j\alpha}\rangle$, where the mean-field parameters $\langle\hat b^{}_{j\alpha}\rangle$ (superfluid order parameters) have to be found selfconsistently. Since the Hamiltonian is a sum over on-site Hamiltonians, the many-body wave function is a product wave function over the lattice sites. However, the different sites are coupled by the superfluid order parameter $\phi_{i\alpha}$. When the bosons are superfluid, the superfluid order parameter is nonzero and the phase is constant in space. This establishes superfluid long range order in the system.

This mean-field approximation is exact in the weak-coupling limit \cite{gutz_U=0}. Moreover, it is exact in the limit of infinite dimensions.  Corrections to the mean-field results scale like $1/z$, where $z$ is the number of neighbors \cite{vollhardt, hubener}. This means that on the cubic lattice, where $z=6$,  the mean-field theory is well-controlled. It provides still qualitatively good results in two dimensions ($z=4$). 

\subsection{Dynamics}

The Gutzwiller mean-field method can straightforwardly be generalized to time-dependent calculations \cite{jaksch}. At zero temperature, the total wave function is assumed to be a product wave function over the lattice sites: $|\Psi(t)\rangle = \prod_i |\psi(t)\rangle_i$, with 
$|\psi(t)\rangle_i=\sum_{n_a, n_b}\beta_{i,n_a,n_b}(t)(\hat b^{\dagger}_{ia})^{n_a}(\hat b^{\dagger}_{ib})^{n_b}|0\rangle$.

Each of the on-site wave-functions is evolved in time according to the local Schr\"odinger equation
\[
i \hbar \partial_t |\psi(t)\rangle_i = \hat{\mathcal{H}}^{MF}_i (t) |\psi(t)\rangle_i \quad.
\]
This constitutes a set of coupled non-linear differential equations for the $\beta_{i,n_a,n_b}(t)$, which have to be solved.
The resulting mean-field dynamics conserves the average total particle numbers $\langle \hat N_\alpha \rangle$. In order to achieve number conservation numerically, the wave function on each lattice site is updated sequentially taking into account the already updated superfluid order parameters of the previous sites. This method keeps the total particle number constant (deviations $<0.1$ \textperthousand) \cite{snoek}, without need of a time-evolving chemical potential \cite{dynamics2}.

\subsection{Finite temperature}

The Gutzwiller mean-field method is also readily generalized to nonzero temperature. In this case the total system is described by a density matrix $\hat\rho(t)$ . Expectation values are obtained as $\langle \hat O \rangle(t) = {\rm Tr}[ \hat \rho(t) \hat O ]$. In mean-\-field  $\hat\rho(t)$ factorizes over the lattice sites as $\hat\rho(t)=\prod_i\hat\rho_i(t)$, with $\hat \rho_i(t) = \sum_n \frac{e^{-\beta E_n^i}}{Z_i} |\psi_n(t) \rangle_i\phantom{}_i \langle \psi_n(t) |$,  where $| \psi_n(t) \rangle_i$ are the local eigenstates of the Hamiltonian $\hat{\mathcal{H}}^{MF}_i(t)$ and $Z_i$ the on-site partition functions. This representation is also valid out of equilibrium. The local density matrix $\hat \rho_i(t)$ obeys the Von Neumann equation
\begin{equation}
i \hbar \partial_t \hat \rho_i(t) = [ \hat{\mathcal{H}}^{MF}_i(t), \hat \rho_i(t)] \quad.
\label{eq:von-neum}
\end{equation}
The von Neumann time evolution is unitary and hence preserves the weights in the density matrix. Thus the thermal weights in the density matrix $\frac{e^{-\beta E_n^i}}{Z_i}$ in the ramp simulation are time-independent, where $E_n$ are the initial eigenvalues of the system and $Z_i$ is the initial partition function. The initial inverse temperature $\beta=1/k_BT$ is set to an experimentally reasonable value. Since in the Gutzwiller approach the density matrix factorizes over the lattice sites, this puts severe restrictions on the possibility of (local) thermalization. In order to account for dissipation, such as induced by three-body losses, one has to apply the Lindblad equation \cite{lindblat}. 

Mean-field theory predicts the existence of long range order at low but nonzero temperatures in any spatial dimension. This means that the correct physics in two-dimensions is not fully recovered at finite temperatures. In three dimensions, mean-field theory describes the correct behavior of the system at zero and nonzero temperature. Our two-dimensional simulations can thus be viewed as a qualitative description of the experiments in \cite{catani}.

\section{Results for $T=0$}
\label{sec:resultsT0}
At low lattice depth the system does not fulfill the single-band and tight-binding approximation required for the Bose-Hubbard Hamiltonian in Eq. (\ref{eq:BH}) to be valid. We therefore start our calculations at $s=5$ where $^{87}$\!Rb is already far in the tight-binding and the lowest band regime. The corresponding lattice depth for $^{41}$\!K is $s_{K}=2.17$. Since the ratio between the bandwidth $4J_K$ and the bandgap $E_{gap}=2\sqrt{s_k}E_r$ is small for the initial $s_K$ ($4J_K/E_{gap}\approx0.03$), the single-band approximation is also satisfied for $^{41}$\!K. We assume that the wave function still corresponds to ground state when the ramp in the experiment reaches this initial lattice depth. This is a valid assumption, since interactions are still weak for $s=5$ and the ramping time in the experiments is chosen adiabatic with respect to the band gap, which guarantees that the particles remain in the lowest band. Starting with $s=4$ and $s=6$ indeed did not change our results.

Although the ramp of the lattice affects both species, only effects of the ramp-up dynamics on the $^{87}$\!Rb-atoms are discussed in the following, like in the experiments \cite{catani}.

To investigate the adiabaticity of the lattice-ramp, we now first present results for the density distribution in real-space as well as in momentum space.  Later on we focus on the visibility.
\begin{figure}[t!]
\includegraphics[width=1.0\linewidth]{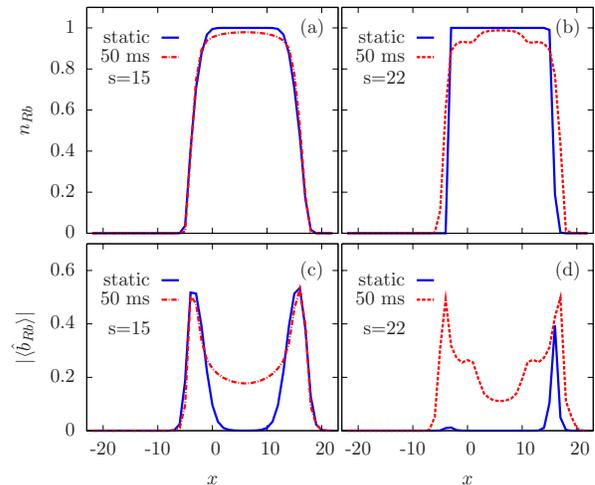}
\caption{(Color online) Particle density and superfluid order parameter of $^{87}$\!Rb along a cut in $x$-direction through the center of the trap ($y=0$). (a) For $s=15$ and ramp duration $t=50$ ms (red dashed line) the particles in the center of the trap are in the ``frozen phase'' in contrast to the ground state profile (blue solid line), where a Mott-plateau is present. This is indicated by the noninteger density $n_{Rb}$ and nonzero local superfluid order parameter $|\langle \hat b_{Rb}\rangle|$ (c). (b) For $s=22$ and $t=50$ (red dashed line) a density wave is induced due to the fast ramp and high nonadiabaticity. 
Global parameters: 
$L=45$, 
$N_{Rb}=303$, 
$N_{K}=30$, 
$U_{Rb-K}=1.93U_{Rb}$. 
Parameters  for (a) and (c):
$J_{Rb}= 0.02U_{Rb}$, 
Parameters for (b) and (d): 
$J_{Rb}=0.004U_{Rb}$.}
\label{fig:na_15_22}
\end{figure}

\subsection{Density profiles}

\subsubsection{Real space} \label{sec:real_space}
To investigate the effect of the ramping dynamics we first compare the density distributions in real-space after the lattice ramp-up with static density profiles at the corresponding final lattice depth. The static profiles originate from the ground state at $T=0$ for a given $s$.  

For small final $s$ ($s<10$) the density profiles agree perfectly with the static ones independently of the ramping time. For larger $s$, longer ramping times are needed to achieve good agreement between the profiles. In particular, for $s\geq15$ in equilibrium a Mott plateau appears in the center of the trap. 
However the dynamically evolved wave function remains a superposition of various Fock states for fast ramps. This can be understood by the following argument. In the limit of an instantaneous increase of the lattice depth the superfluid wave function consisting of local superpositions of various Fock states remains unchanged. Since the hopping $J$ decreases exponentially with increasing lattice depth, after this sudden step to a deep lattice the time-evolution operator consists mainly out of the interaction part $U\hat n(\hat n-1)/2$ which rotates the phase of each Fock state independently, but leaves the local wavefunction in a superposition of multiple Fock states.
The situation is similar for fast ramp and leads to noninteger particle density (see Fig.~\ref{fig:na_15_22}(a)(b)) and nonzero local superfluid order parameter (see Fig.~\ref{fig:na_15_22}(c)(d)). Although the absolute value of the superfluid parameter is not vanishing, long-range order is destroyed as the complex phases of $\langle \hat b_i\rangle$ are not constant over the lattice any more. Correlations between the phases, however, can be partially recovered for certain ramp-up times. This corresponds to the collapse-and-revival physics \cite{bloch-osci}: after time intervals of length $2 \pi \hbar/U$ all the individual phases are back in phase and global phase coherence is restored. Away from the revival times, due to the vanishing coherence this phase is not a SF-phase, we refer to it as a ``frozen'' phase.  

Ramping the lattice to higher $s$ with the same ramp-up time means effectively faster increase of the lattice depth per time unit. This causes not only significant deviations from the static density profiles but also the formation of density waves (Fig.~\ref{fig:na_15_22}(b)). This behavior is observed for ramp-up times below 100 ms. The density waves are a further evidence that the final state reached after the lattice ramp is not necessarily the ground state.

We finally examine the rotational symmetry of the density profile of the Rb atoms.
Due to the presence of the K species (and hence additional repulsive interactions), the rotational symmetry of the Rb cloud is always broken. However, this effect is only very pronounced for large $s$. Deep in the MI regime  the compressible superfluid boundary layer, adjacent to the cloud of K-atoms, disappears because of the strong repulsive force experienced by the K atoms. The corresponding superfluid order parameter vanishes in the ground state.
This is different in the case of the time-evolved profiles: as the $^{87}$\!Rb atoms for $t=50$ $s=22$ ms are in the ``frozen'' phase, the breaking of circular symmetry is far less pronounced as in the static case, where the system is already Mott-insulating.

\begin{figure}[t!]
\includegraphics[width=1.0\linewidth]{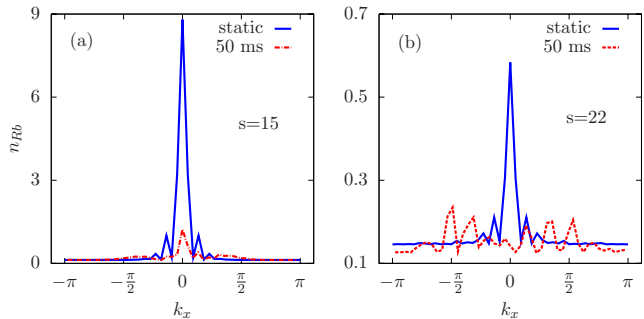}
\caption{(Color online) Particle density in momentum space along a cut in (1,0)-direction for $k_y=0$. (a) In the static calculations (blue solid line) the central peak at $\vec{k}=(0,0)$ is present, due to the SF shell, and small side peaks at nonzero $\vec{k}$ due to the MI-core in the middle of the trap. Fast ramps with $t=50$ ms (red dashed line) drive the particles into a frozen phase with lowered phase coherence, which is mirrored in a decreased central peak. (b) Ramping within the same time to a deeper lattice is highly nonadiabatic. The central peak becomes smaller than the side peaks or even the background value $n(\vec k = (\pi, \pi))$ and the $\pm \vec{k}$ symmetry disappears.}
\label{fig:nk_15_22}
\end{figure}

\subsubsection{Momentum space}
More information regarding superfluid long-range order is available in momentum space. Besides, in contrast to the real-space particle distribution, information on the momentum distribution is experimentally well accessible by time-of-flight measurements. The nonadiabaticity and oscillations in the density profile are mirrored in the momentum distribution. To accentuate the global behavior as a function of the ramp-up time and the final lattice depth we show only the values averaged over an equilibration time of 16 ms.
 
Whereas for a homogeneous system of noninteracting particles the momentum distribution is a single delta-peak at $ \vec k=(0,0)$,  the momentum distribution of a trapped system of interacting particles has a constant background due to non-condensed particles and shows broadening because of the trap. This latter effect dominates over the broadening by short-range fluctuations, which are neglected in the Gutzwiller approximation.

As pointed out in the previous subsection, for short ramp-up times and deep lattices the local superfluid order parameter remains finite in the region where the static calculation predicts a MI-plateau. Particles stay ``frozen''  in a superfluid-like phase indicated by noninteger particle densities and finite local superfluid order parameter. Although in the dynamic case all Rb atoms seem to be superfluid, the phase coherence is lost. This is shown in Fig.~\ref{fig:nk_15_22}. The reduced central peak in the $n(\vec k)$-profile compared to the static data clearly indicates destroyed long-range order in the system (Fig.~\ref{fig:nk_15_22}(a)). In extremely nonadiabatic cases the central peak becomes smaller than the side peaks or even the background value $n(\vec{k}=(\pi,\pi))$. This leads to a significant broadening of the momentum distribution. Additionally the $\pm \vec{k}$ symmetry  breaks down (Fig.~\ref{fig:nk_15_22}(b)).

\subsection{Visibility} \label{sec:visibility}
The most convenient way to compare the experimental momentum distribution with our theoretical results is to calculate the visibility $\eta$ for the $^{87}$\!Rb atoms as a function of lattice depth $s$
  \begin{equation}
	\eta(s) = \frac{n_s(\vec{k}=(0,0))-n_s(\vec{k}=(\pi,\pi))}{n_s(\vec{k}=(0,0))+n_s(\vec{k}=(\pi,\pi))}\quad.
\label{eq:vis}
\end{equation}
Here $n_s(\vec{k})$ corresponds to the spatial Fourier transform of $\langle\hat{b}^{\dagger}_i \hat{b}^{\phantom{a}}_j\rangle(s)$. In the experimental procedure \cite{catani} the height of the first order peaks is compared with the minimum in $n(\vec{k})$ at the same distance from the central zero order peak, to divide out the contribution of the Wannier function. Our calculations are performed within the tight-binding model Eq. (\ref{eq:BH}) and do not include the shape of the Wannier function. We therefore calculate the visibility by comparing the central peak with the minimum at the edge of the Brillouin-zone.

The $n(\vec{k})$ values are extracted in the experiment \cite{catani} by integrating finite square areas around the peaks instead of taking single values. Application of this method to our theoretical data shifts the visibility for all ramp-up times. This is a small quantitative effect and depends on the extent of the integration area. It has no effect on the conclusions drawn in this paper.

In order to investigate the reduction of the visibility in a systematic way, we now subsequently analyze the role of the second species, the ramp-up profile of the lattice and the ramp-up time.

\subsubsection{Effect of the second species} 

\begin{figure}
\includegraphics[width=1.0\linewidth]{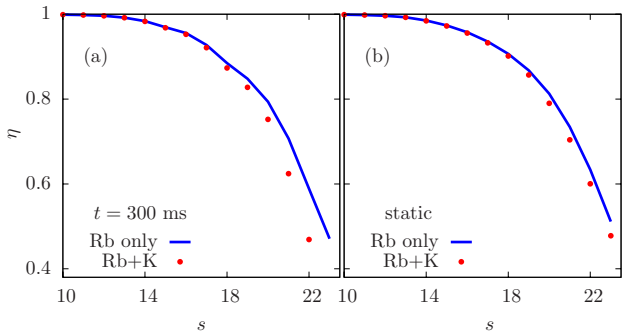}
\caption{(Color online) Visibility for pure $^{87}$\!Rb (blue solid line) and in the presence of $^{41}$\!K (red dots) after a $t=300$ ms ramp (a) and in the static case (b). In both cases the visibility of pure $^{87}$\!Rb is higher.}
\label{fig:vis_lin+-K}
\end{figure}

It is observed experimentally that addition of  $^{41}$\!K to a system of  $^{87}$\!Rb-particles reduces the phase coherence and the visibility \cite{catani}. 
Our simulations reproduce this behavior for similar parameters and particle number ratios as in the experiment, i.e. $N_{Rb}\approx 303$ and $N_K \approx 30$. We observed this effect for all simulated ramping times (from $50$ to $300$ ms with a linear ramp-up profile) as well as in the static case (Fig.~\ref{fig:vis_lin+-K} (a),(b)). In particular, the destructive effect of the $^{41}$\!K on the phase correlations of the superfluid order parameter of $^{87}$\!Rb is more pronounced in the  dynamic case than in the static one. This indicates that the second species enhances the nonequilibrium induced by the lattice ramp. 

The reason for the lower $^{87}$\!Rb visibility in the presence of $^{41}$\!K in the static case is the following. The repulsive interaction between the species pushes the $^{87}$\!Rb atoms out of the overlap region and enhances the $^{87}$\!Rb density in the trap center. This increases the interaction energy of the $^{87}$\!Rb and brings the particles closer to the SF-MI transition, thus decreasing the coherence of the system in the static case \cite{buonsante}. 

The dynamic lattice ramp to deep lattices leads even to a lower visibility than in the static case. This can be explained by the following argument.
As we will show below, one of the reasons for the nonadiabaticity of the ramp-up is the excitation of collective modes. This effect results from the interplay between the intraspecies repulsion and the increase of the lattice depth, which squeezes the atomic cloud. 
When $^{41}$\!K is present,  the clouds exert a repulsive force on each other, even though the overlap region is small. This leads to additional collective modes in the system, which lowers the phase coherence and hence the visibility. Below we will quantify this statement.
 This mechanism could explain why in experiments always a reduced visibility is seen when a second species is added.

\begin{figure}
\includegraphics[width=1.0\linewidth]{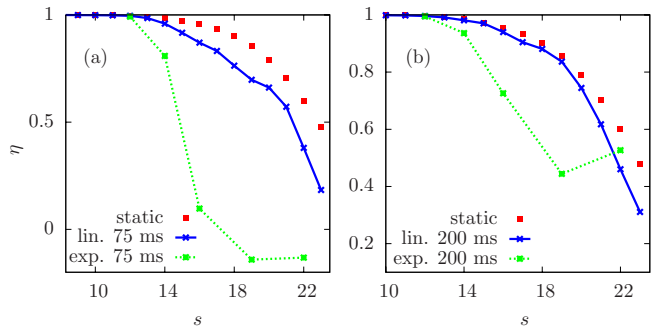}
\caption{(Color online) Effect of the ramping profile on the visibility of $^{87}$\!Rb in the presence of $^{41}$\!K. (a) The resulting visibility after $t=75$ ms ramp with the linear ramp (blue solid line) lies closer to the adiabatic static visibility (red solid line) than the visibility after ramp with the exponential profile (green dashed line). (b) Results with the same profiles but for $t=200$ ms ramp.}
\label{fig:vis_lin+exp}
\end{figure}

\subsubsection{Effect of the ramp-up profile}
In this subsection we investigate the effect of the time profile of the lattice ramp on adiabaticity. Our motivation is that in experiments the ramping profiles are usually of exponential shape to keep particles in the lowest band at the beginning of the ramp. For a better comparison between the simulated linear profile and the one used in experiment we also used the same shape as in \cite{catani} 
\begin{equation}
	s^\prime(t^\prime)=( e^{\frac{t^\prime}{0.4t}} - 1 )\frac{s}{ e^{\frac{1}{0.4}} - 1 }\quad,
	\label{eq:ex-profile}
\end{equation}
where $s$ is the final lattice depth and $t$ the ramping time. We performed calculations for $50$to $300$ ms ramp-up times. These simulations demonstrate that the exponential ramp leads to a lower visibility than the linear profile (Fig.~\ref{fig:vis_lin+exp}). The highly nonadiabatic ramp-up time of 75 ms with exponential profile leads even to negative visibilities (Fig.~\ref{fig:vis_lin+exp}(a)). This indicates that the system displays collapse-and-revival physics for these short ramping times. Extending the ramping time to $t=200$ ms results in an almost adiabatic linear ramp for $s\leq15$ as the calculated visibility corresponds very well to the static (Fig.~\ref{fig:vis_lin+exp}(a)). The visibility of $^{87}$\!Rb after $200$ ms exponential ramp is higher than after $75$ ms but still lower than the linear results. For deep lattices with $s=22$ the exponential ramp-up profile leads to a higher visibility than the linear ramp. However, this is due to the fact that the exponential ramp-up is still highly non-adiabatic for this ramping time, leading to a large SF-fraction, whereas the linear ramp is closer to being adiabatic.   

The observation that the exponential ramp-up profile leads to less adiabatic behavior is explained by the fact that this profile has a low ramp-up velocity for the smaller lattice-depths, where the interaction only plays a minor role, whereas for the higher lattice depths, where many-body effects become important, the ramping velocity is very high. 
For this reason further investigations were performed exclusively with a linear ramp-up shape. 

\begin{figure}[top]
\includegraphics[width=1.0\linewidth]{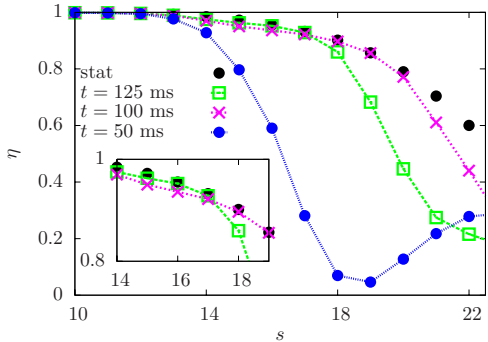}
\caption{(Color online) Calculated visibilities for $125$ ms (green squares) $100$ ms (pink crosses) and $50$ ms (blue dots) ramping durations compared to the static visibility (black dots, not connected by line).  The inset demonstrates that for $s<17$ the visibility after a 125 ms ramp lies closer to the static visibility than after a 100 ms and vice versa for deeper lattices. The increase of the visibility for deep lattices for $t=50$ ms is due to the highly non-adiabatic revival of the coherence.}
\label{fig:vis_allt}
\end{figure}

\subsubsection{Effect of the ramp-up time}
\begin{figure}[t]
\includegraphics[width=1.0\linewidth]{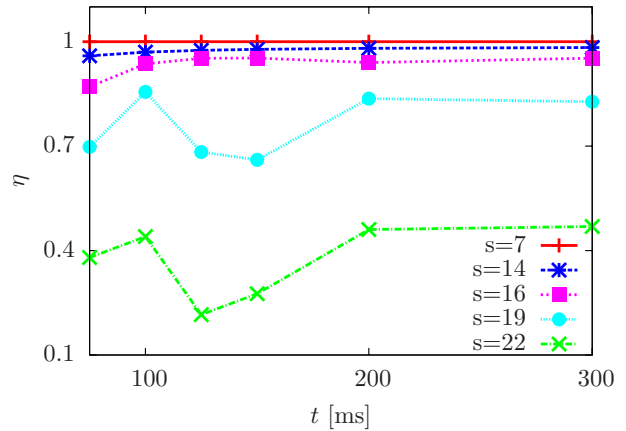}
\caption{(Color online) Visibility as a function of ramp-up time. For $s=7$ (red solid line) all considered ramping times are adiabatic and the visibility is independent of the ramp-up time. For $s=14$ (blue asterisks) longer ramp leads to the ground state visibility, indicating $300$ ms as the appropriate ramp-up time. For $s\geq17$ oscillations are induced, as explained in the text below.}
\label{fig:vis_alls}
\end{figure}

Finally we investigate the effect of the ramp-up time  on the visibility. Whereas one would expect that a slower ramp automatically enhances the adiabaticity and the visibility,  in the simulations we observe that only the dynamic real-space particle distributions become similar to the static ones. The visibility, and therefore the phase coherence, remains different from the ground state one. 
Especially for intermediate ramping times the behavior is counterintuitive (Fig.~\ref{fig:vis_allt}).  While for lattice depth $s<15$ the visibility increases monotonically  with the ramp-up time (Fig.~\ref{fig:vis_allt} inset), deeper in the MI-regime this tendency is washed out. This is surprising as the longer ramp is expected to be more adiabatic. Plotting the visibility as a function of ramp-up time for a fixed $s$ demonstrates that the effect of the ramping time is not the same for all $s$ (Fig.~\ref{fig:vis_alls}). We distinguish two different regimes. Ramping up to shallow lattices $s\leq12$ ($\hbar/J(s=12)\approx5.4$ ms) is adiabatic on all time scales and thus not affected by the ramp duration. In the regime of final lattice depth around the SF-MI transition and in MI-phase, oscillations  in the visibility occur.

\begin{figure}[t]
\includegraphics[width=1.0\linewidth]{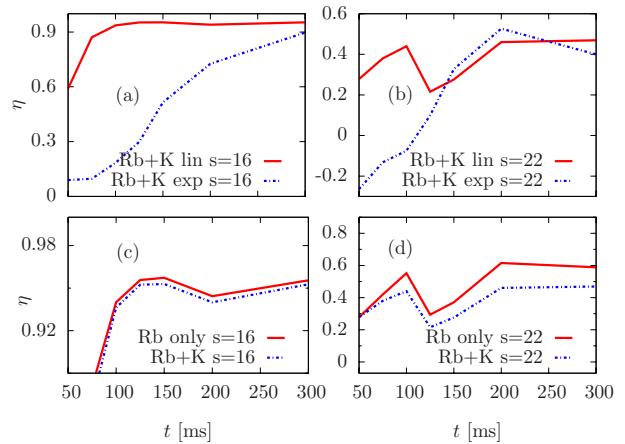}
\caption{(Color online) Effect of the ramping profile on the visibility oscillations (a), (b) and effect of the second species (c), (d). For intermediate lattices e.g. $s=16$ the exponential ramp-up profile (blue dashed line) damps the oscillations, which are present for the linear profile (red solid line) (a). For $s=22$ and exponential ramp (b) a non-adiabatic maximum appears. The presence of a second species (blue dashed line) (c), (d) only induces a global shift of the visibility.}
\label{fig:osci_ramp_Rb+-K}
\end{figure}

These oscillations were also observed experimentally \cite{catani, bloch-osci} but not yet explained. In the next subsection we give an explanation in terms of coupling to the collective modes of the system. Before this we add some remarks on the figures. The oscillating behavior depends on the ramp-up profile (Fig.~\ref{fig:osci_ramp_Rb+-K}(a)(b)). The exponential shape shifts the oscillations to longer ramp-up times. The presence of $^{41}$K has only a minor effect on the oscillations: the second species only leads to a global shift in the visibility without changing the position of maxima and minima (Fig.~\ref{fig:osci_ramp_Rb+-K}(c)(d)). This is consistent with the experimental findings \cite{catani}.

\begin{figure}[tb]
\includegraphics[width=1.0\linewidth]{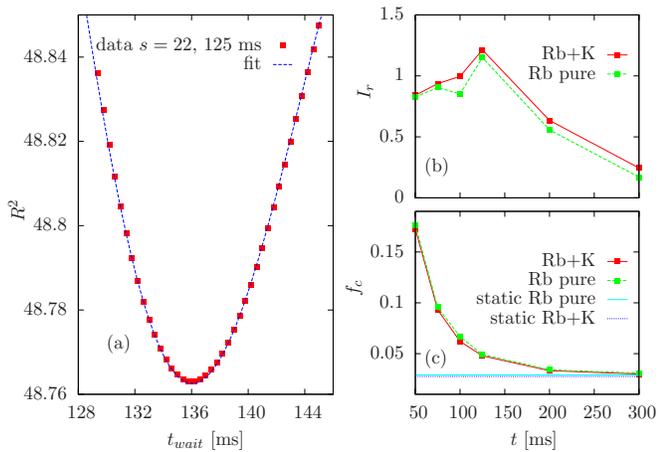}
\caption{(Color online) (a) Oscillations of the radius of the Rb-cloud during the waiting time of 16 ms after a $t_0= 125$ ms ramp to $s=22$. The data is fitted by $f(t)=0.135\sin(0.18t+5.19)\exp^{-0.031(t-t_0)}+48.857$ (blue line). The sinusoidal form corresponds to the excitation of the breathing mode. (b) The current at $s=22$, averaged over the waiting time and renormalized with respect to SF-fraction, induced after different ramp-up times. The maximum of the renormalized current indicates the regime with the maximal coupling to the collective modes. The second species leads to an enhanced coupling for all ramp-up times. (c) The SF-fraction averaged over the waiting time as a function of ramping time at $s=22$. The ``freezing'' in the SF-phase is dominant for fast ramps. For longer ramp-up the system approaches the ground state (static) value and the ramp is getting more adiabatic. The $^{41}$\!K slightly reduces the SF-fraction.}
\label{fig:multiplot_ampl_x2}
\end{figure}

\subsubsection{Explanation of visibility oscillations}
We now turn to the explanation of the oscillations in the visibility. A closer look at the simulations shows that the only component contributing to the visibility which oscillates in time at a fixed $s$ is $n(\vec{k}=(0,0))$. The particle density in momentum space is given by
\begin{equation}
 n(\vec{k}=(0,0))=\frac{1}{L^2}\sum_{i=1}^{L^2}\bigg(\langle\hat n_i\rangle-r_i^2\bigg)+\frac{1}{L^2}\bigg|\sum_{i=1}^{L^2}r_i\mathrm{e}^{-i\phi_i}\bigg|^2 \,, 
\end{equation}
where $L^2$ is the total number of lattice sites,  $r_i (r_j)$ and $\phi_i(\phi_j)$ are the absolute value and the phase of the local superfluid order parameter on a site $i$ ($j$) respectively: $\langle b_i^\dagger \rangle = r_i e^{-i \phi_i}$. We can separate the contribution of the absolute value and the phase to the visibility. The absolute value $r_i$ shows monotonic behavior and is continuously decreasing if the ramp-up time is reduced. This is measured by the Gutzwiller SF-fraction $f_c = \sum_i |\langle b_i \rangle|^2/N$. In Fig.~\ref{fig:multiplot_ampl_x2}(d) the SF-fraction is continuously decreasing and always higher than the equilibrium value. Only for long ramping times ($t = 300$ ms), the dynamical and static SF-fraction approach each other. This means that for the short ramping times the visibility is dominated by this anomalously large SF-fraction. 
However, for very fast ramping times the SF-fraction is high, but the phases are uncorrelated, leading to a low visibility. With increasing ramping time the phase coherence builds up and compensates the decay of the SF-fraction leading to an increasingly higher visibility. 

For even longer ramping times the visibility decreases again.
 We explain this by an enhanced coupling to the collective breathing mode of the system 
induced by the lattice ramp. Increasing the lattice depth results in an increased ratio of the strength of the harmonic trapping potential and the hopping constant $J$. This forces the particles to move towards the center of the parabolic trap and leads to a higher occupancy in the middle of the trap. At the same time the particles experience a higher repulsion as $U(s)$ grows with the lattice depth (see Eq. (\ref{eq:param-formel})). This repulsion acts against the increasing population and induces a reverse flow. The interplay between these two mechanisms yields the collective modes.

We indeed find numerical evidence for these collective oscillations by observing the  cloud size $R^2=\langle \vec r^2\rangle - \langle \vec r\rangle ^2$ during the waiting time after the ramp. The sinusoidal oscillations of $R^2$ indicate the collective movement of particles within the breathing mode in Fig.~\ref{fig:multiplot_ampl_x2}(a). The data for other ramp-up times can also be fitted similarly. This leads to the conclusion that only the breathing mode is excited. Depending on the ramping time the amplitude of the oscillation changes. In particular, the amplitude is continuously decreasing when the ramping time is made longer. However, this is mainly due to the decreasing SF-fraction, which reduces the number of mobile particles. Renormalizing the amplitude by the SF-fraction leads to a peak at the position of the minimum of the visibility, which evidences that the coupling to the modes is responsible for this minimum. 

Analysis of the total current in the system $I=\sum_{\langle ij\rangle} | \langle \hat b^\dagger_i \hat b_j - \hat b^\dagger_j \hat b_i\rangle|^2$ further clarifies this. The total current is decaying because of the decaying SF-fraction. To investigate the relative motion of the mobile particles we therefore renormalize the total current by $f_c^2$: $I_r=I/Nf_c^2$. This function again shows a clear maximum at the position of the visibility minimum (see Fig.~\ref{fig:multiplot_ampl_x2}(c)). As the coupling to collective modes destroys the phase coherence, this fully agrees with the minimum in visibility for a 125 ms ramp, see Fig.~\ref{fig:osci_ramp_Rb+-K}(d). 

For $t>150$ ms the SF-fraction approximates the static value and the collective modes are less excited. This is the most adiabatic ramping regime. 

For the exponential ramp-up profile, the oscillations are shifted (Fig.~\ref{fig:osci_ramp_Rb+-K}(a)(b)). This is because the SF-fraction remains anomalously high even for long ramping times and the visibility is dominated by this effect. 
In particular the maximum at $t_{\rm ramp} = 200$ ms for $s=22$ is explained by the high SF-fraction and is thus a highly non-adiabatic point. The decrease of the visibility at $t=300$ ms for $s=22$ is explained by an enhanced coupling to collective excitations.

The fact that the presence of a second species leads only to a small shift in the visibility is explained by the observation that the overlap of the atomic clouds is very small and the modes are mainly excited by the increased repulsion between the $^{87}$Rb particles when the optical lattice is ramped up.
However, the second species induces additional modes, which lower the visibility. This is seen in the higher renormalized current in Fig. \ref{fig:multiplot_ampl_x2}(b) for the mixture compared to the single-species system.

The additional induced modes in the system due to the presence of $^{41}$\!K not only explain the lower visibility, but also offer an explanation for the experimental observation that adding a second species leads to a broadening of the momentum profile beyond a certain lattice depth \cite{catani}. The presence of collective modes leads to macroscopic occupation of single particle states with nonzero momentum and hence broadens the momentum distribution. It is worth noting that for this explanation the amount of spatial overlap of the two species is less important: as long as the two clouds touch, they can exert a force on each other. This explains why this effect was already found for widely separated components.

\section{Results for nonzero T}
\label{sec:resultsT}
In order to understand the effect of finite temperature in the experiments we also perform simulations for this case. We perform simulations for initial temperatures of $19$ nK ($k_BT_2=2.2J_{Rb,s=5}$) and $12.6$ nK ($k_BT_1=1.5J_{Rb,s=5}$). This is in the range of typical experimental temperatures, which can be estimated as detailed in appendix \ref{sec:appendix}.
To investigate the adiabaticity of the ramp at finite temperature we again compare time-dependent ramp-up simulations with static results. The latter correspond to the ensemble in thermal equilibrium at a final lattice depth with effective inverse temperature $\tilde\beta=1/k_B\tilde T$.  The effective temperature is chosen such that the entropy of the static system equals the initial entropy of the ramped system. The static results thus represent an adiabatically ramped and completely thermalized ensemble.
Accordingly, the static density matrix is $\rho=\sum_n\frac{e^{-\tilde\beta E_n^i}}{Z_i} |E_n\rangle_i\phantom{}_i \langle E_n |$, where $E_n, |E_n\rangle_i$ are the eigenvalues and eigenstates of the Hamiltonian at site $i$ respectively. 
\begin{figure}[t!]
\includegraphics[width=1.0\linewidth]{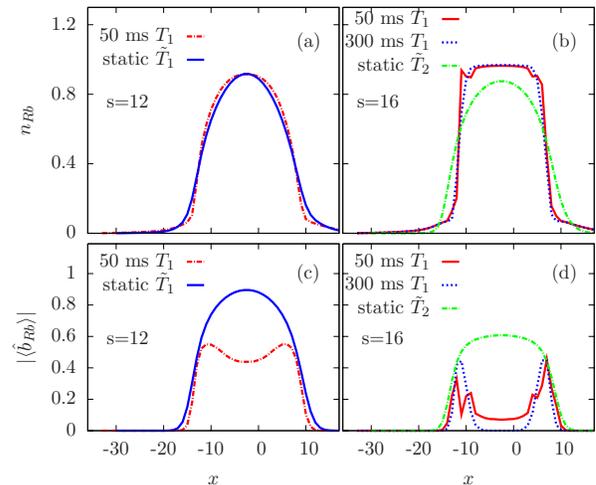}
\caption{(Color online) Finite temperature results for the particle density and superfluid order parameter of $^{87}$\!Rb at final lattice depth $s=12, 16$ along a cut in $x$-direction through the center of the trap $(y=0)$. All dynamic profiles are calculated for the temperature $k_BT_1=1.5J_{Rb,s=5}$. (a) For $50$ ms ramp (red dashed line) the results correspond well to the static thermalized ensemble (blue solid line) with effective temperature $k_B\tilde T_1=0.95J_{Rb,s=5}$. However, the local superfluid order parameter is reduced in the center of the atomic cloud (c). (b) After the ramp to $s=16$ the dynamic profiles differ from the static ones with corresponding effective temperature $k_B\tilde T_2=0.42J_{Rb,s=5}$. A density plateau is formed in the center of the trap at a noninteger density. For 50 ms ramp (red solid line) density waves appear around this plateau. (d)
The superfluid order parameter (red solid line) is peaked in the region where the density waves appear.
Parameters: 
$L=60$, 
$N_{Rb}=303$, 
$N_{K}=30$,  
$U_{Rb-K}=1.93U_{Rb}$,
Parameters for (a) and (c):
$J_{Rb}=0.24U_{Rb}$,
Parameters for (b) and (d):
$J_{Rb}=0.02U_{Rb}$.} 
\label{fig:na_12,16_beta0.6}
\end{figure}
\subsection{Density profiles}
We first investigate the density profiles in real-space at $k_BT_1=1.5J_{Rb,s=5}$. Ramping in 50 ms to $s=12$  provides a density profile similar to the static thermalized result with effective temperature $k_B\tilde T_1=0.95J_{Rb,s=5}$ (see Fig.~\ref{fig:na_12,16_beta0.6}(a)). The local superfluid order parameter, however, differs from the value in thermal equilibrium (see Fig.~\ref{fig:na_12,16_beta0.6}(c)). In the center of the $^{87}$\!Rb cloud the local superfluid order parameter is reduced. Hence the system after the ramp does not correspond to an adiabatically ramped and thermalized ensemble. At deeper lattices the dynamic density profiles do not fit the thermal static distribution. Fig.~\ref{fig:na_12,16_beta0.6}(b) demonstrates the squeezing of the dynamic profile for $s=16$ compared to the static one with effective temperature $k_B\tilde T_2=0.42J_{Rb,s=5}$. Similar as in the $T=0$ case, the fast 50 ms ramp is nonadiabatic and induces density waves around the plateau (see Fig.~\ref{fig:na_12,16_beta0.6}(b)). The peaked local superfluid order parameter in Fig.~\ref{fig:na_12,16_beta0.6}(d) corresponds to the region where the density waves appear. For $t=300$ ms in Fig.~\ref{fig:na_12,16_beta0.6}(b) a plateau is formed at non-integer density in the center of the trap. At the same time the corresponding local superfluid order parameters vanish which indicates a formation of a normal phase instead of a MI-plateau as found previously in the $T=0$ case.  
\begin{figure}[t]
\includegraphics[width=1\linewidth]{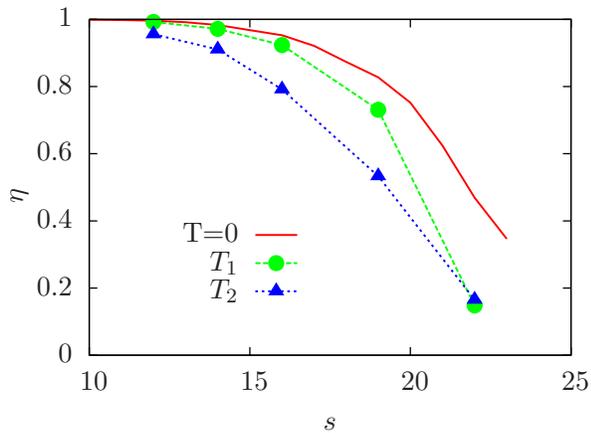}
\caption{(Color online) Visibility after 300 ms ramp-up for $k_BT_1=1.5J_{Rb,s=5}$ (green dots) and $k_BT_2=2.2J_{Rb,s=5}$ (blue triangles) compared with $T=0$ (red solid line). The visibility decreases with increasing temperature.}
\label{fig:vis_allbeta_t=300}
\end{figure}
\begin{figure}[t]
\includegraphics[width=1\linewidth]{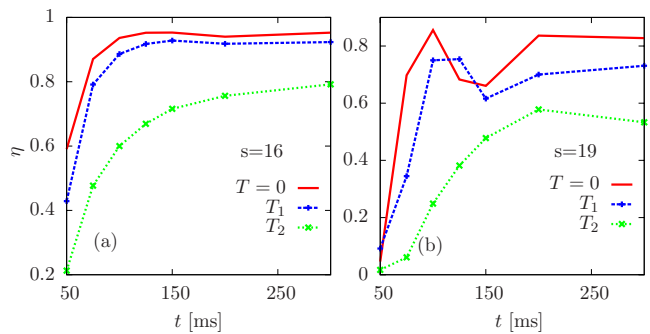}
\caption{(Color online) Visibility as a function of ramp-up time for different temperatures and final lattice depths. (a) At shallow lattices for $T=0$ (red solid line) oscillations appear with a visibility minimum at $t=200$ ms. For $k_BT_1=1.5J_{Rb,s=5}$ (blue dashed line) the global visibility and the oscillation amplitude is reduced. Higher temperature ($k_BT_2=2.2J_{Rb,s=5}$, green dotted line) blurs the oscillations. (b) For $s=19$ the shape of the oscillations at $T=0$ is mainly conserved for $k_BT_1=1.5J_{Rb,s=5}$. The graph is shifted to lower visibility and longer ramp-up times. At $k_BT_2=2.2J_{Rb,s=5}$ only one maximum at $t=200$ ms is present.}
\label{fig:vis-osci-allbeta}
\end{figure}

\subsection{Visibility and Oscillations}
As shown in the previous subsection, with increasing temperature and lattice depth the condensate depletes. This also lowers the visibility. In Fig.~\ref{fig:vis_allbeta_t=300} this behavior is exemplified for a $300$ ms ramp. For low temperature $k_BT_1=1.5J_{Rb,s=5}$ and shallow lattices $s\leq12$ the visibility is hardly changed compared to $T=0$. With increasing lattice depth, however, the superfluid order parameter vanishes in the trap center, leading to a decreased visibility. For higher temperature ($k_BT_2=2.2J_{Rb,s=5}$) the fraction of the atoms in the normal phase increases, thus lowering the visibility further.  

The visibility oscillations are affected as well. When temperature and final lattice depth are sufficiently small, the only effect of the temperature is to reduce the visibility (Fig.~\ref{fig:vis-osci-allbeta}(a) $k_BT_1=1.5J_{Rb,s=5}$). Although the absolute value of the local superfluid parameter is reduced in the center of the trap for slow lattice ramp compared to fast ramp (Fig.~\ref{fig:na_12,16_beta0.6}(d)), the phase coherence in the latter case is almost completely destroyed. This leads to a higher visibility for a $300$ ms ramp than for a $50$ ms.  For $k_BT_2=2.2J_{Rb,s=5}$ the visibility oscillations are suppressed and the minimum at $t=200$ ms disappears. 
We can understand the disappearance of the minimum in the visibility in a qualitative way by comparing the excitation energy of the breathing mode with the temperature. We indeed find that the temperature here is higher than the excitation energy, meaning that the mode is already thermally occupied and that the ramp of the lattice has less effect.

\section{Discussion and conclusion}\label{sec:conclusions}
Using a time-dependent Gutzwiller model for an interacting Bose-Bose mixture we investigated the ramp-up of the optical lattice for zero and at finite temperatures.

The non-adiabaticity of the lattice ramp was analyzed by comparing density profiles in real and momentum space and by studying the visibility. The adiabatic regime is reached when the density profiles as well as visibility agree with the equilibrium results.

We have shown that a ramp-up of the optical lattice carried out on a time scale comparable to the tunneling time does not necessarily provide the ground state of the system. Depending on the ramp-up time, ramping the lattice at $T=0$ leads to the trapping of the particles in a "frozen" phase with non-integer particle number and nonzero local superfluid order parameter but vanishing global phase coherence. Ramping the lattice at finite temperature additionally causes a depletion of the condensate. The latter grows with increasing temperature. Both ramp effects lead to decreased visibility. 

The ground state visibility was only reached for shallow lattices within the investigated ramping times. The fact that one needs rather long times to be completely adiabatic for deep lattices, is rooted in the critical slowing down of the hopping at the SF-MI-phase boundary at $T=0$ or the SF-normal-phase at finite temperature.

For $T=0$ we demonstrated in addition that the linear lattice ramp is more adiabatic than the exponential. We found that a longer ramp-up time does not naturally lead to a better visibility. In fact, depending on the final lattice depth oscillations may occur for $T=0$ and low temperatures.

We explain these oscillations by a coupling of the ramp-up process to the collective modes of the system. This is consistent with the appearance of density waves in the system. Lowered superfluidity and a larger normal phase prevents collective excitations at higher temperatures.  
One of our main results is that the maximum in the visibility is not a good indication of adiabaticity. This regime is in fact highly non-adiabatic, since the maximum is caused by an anomalously large SF-fraction induced by the short ramping time.  
The region where the visibility saturates is the most adiabatic. However in our approach three body collisions and heating is not included. These processes become relevant at long time scales and also lead to non-adiabaticity.

In experiments, the presence of a second species destroys the phase coherence of the majority species, leading to a decreased visibility. In contrary, previous theoretical static calculations predict either enhanced or decreased long range order, depending on the actual particle ratio \cite{buonsante}. Here we observed that for $T=0$ the dynamical ramp induces additional non-equilibrium and leads to a more pronounced visibility decay  in the region where static calculation also predicts lowering. We explained this in terms of an enhanced current in the system.
This supplementary visibility reduction could be the reason for the experimental findings.

\begin{acknowledgments}
The authors acknowledge informative discussions with  F. Minardi and G. Modugno. This work was supported by the Deutsche Forschungsgemeinschaft (DFG) via Forschergruppe FOR 801 and by the Nederlandse Organisatie voor Wetenschappelijk Onderzoek (NWO). 
\end{acknowledgments}	

\appendix
  \section{Estimation of the Temperature in a weak lattice} \label{sec:appendix}
The ramp-up in our work starts at $s=5$ and not at \mbox{$s=0$}, as in the experiment, due to the required tight-binding regime of the Hubbard Model. Therefore, starting from $s=5$ we have to recalculate the temperature based on the initial experimental temperature $73$ nK \cite{catani} before the ramp. It can be shown by the following argument that the ramp-up of the optical lattice cools the system down. For a first estimate let us assume that the lattice is ramped up adiabatically from $s=0$ to $s=5$ and that the particles are noninteracting. The initial slow ramp of the exponential or any other ramping profile realizes the first condition. The second condition is strictly satisfied only in shallow lattices. For $s=5$, where $U_{Rb}\sim J_{Rb}$, it can be assumed to be roughly satisfied and leads to the right temperature range.

In the case of an adiabatic lattice ramp the entropy of the system $S=-k_B\sum_k\bigg(\ln[1+n_k](1+n_k)-n_k\ln n_k \bigg)$ remains constant. As the experiments start from $s=0$, the initial dispersion corresponds to the free particle dispersion $\varepsilon_k^i=\hbar^2\vec k^2/2m$. The final situation is a tight-binding case and the particle density is evaluated with the dispersion $\varepsilon_k^f=-2J(\cos(k_xa)+\cos(k_ya))$, where $a$ is the lattice constant. As the noninteracting condensate is located around $\vec k=(0,0)$ the tight-binding dispersion can be approximated by a Taylor series as $\varepsilon_k^f\approx J(\vec k a)^2$. The initial and final particle densities $n_k^{i,f}=(e^{\beta\varepsilon_k^{i,f}}-1)^{-1}$ have the same functional dependence on momentum $k$. The sum over functions with the same functional $k$-dependence in entropies can only remain constant when these functions are identical. This is realized when the particle density $n_k$ for the initial lattice depth corresponds to the final one. From the equality of initial and final particle densities the criterion for the final temperature follows $T^f/T^i=\varepsilon^f_k/\varepsilon^i_k$. From here the final temperature can be estimated $T^f=T^iJ\lambda^2m/2\hbar^2$, where $\lambda$ is the laser wave length and $J$ is the hopping amplitude at the final lattice depth \cite{adiab_load}. 

For a lattice ramp to $s=5$ the initial temperature of \mbox{$73$ nK} is lowered by a factor $2$. As for this lattice depth the system is in the SF regime, the thermal energy competes with the hopping energy $J$. At this point the dimensionality of the experiment has to be taken into account. As the hopping scales with the number $z$ of the next neighbors the ratio $T^f/zJ$ for our 2D system studied here should correspond to the ratio of the experimental 3D system. This lowers the calculated temperature by an additional factor $2/3$, providing the effective temperature $T^f=24$ nK for our equivalent 2D system.


\end{document}